\documentclass{optica-article}

\articletype{Research Article}

\usepackage{lineno}
\usepackage[separate-uncertainty=true]{siunitx}
\begin{document}

\title{Terahertz-Induced Nonlinear Response in ZnTe 
}

\author{Felix Selz,\authormark{1,2,3,*} Johanna Kölbel,\authormark{2} Felix Paries,\authormark{1,3} Georg von Freymann,\authormark{1,3} Daniel Molter,\authormark{1} and Daniel M. Mittleman\authormark{2}}

\address{\authormark{1}Fraunhofer Institute for Industrial Mathematics ITWM, Department Materials Characterization and Testing, 67663 Kaiserslautern, Germany\\
\authormark{2}School of Engineering, Brown University, Providence, Rhode Island 02912, USA\\
\authormark{3}Department of Physics and Research Center OPTIMAS, RPTU Kaiserslautern-Landau, 67663 Kaiserslautern, Germany}

\email{\authormark{*}felix.selz@web.de}

\begin{abstract*} 
Measuring terahertz waveforms in terahertz spectroscopy often relies on electro-optic sampling employing a ZnTe crystal. Although the nonlinearities in such zincblende semiconductors induced by intense terahertz pulses have been studied at optical frequencies, a quantitative study of nonlinearities in the terahertz regime has not been reported.
In this work, we investigate the nonlinear response of ZnTe in the terahertz frequency region utilizing time-resolved terahertz-pump terahertz-probe spectroscopy. 
We find that the interaction of two co-propagating terahertz pulses in ZnTe leads to a nonlinear polarization change which modifies the electro-optic response of the medium at terahertz frequencies.
We present a model for this polarization that showcases the second-order nonlinear behavior. We also determine the magnitude of the third-order susceptibility in ZnTe at terahertz frequencies, $\chi^{\mathrm{(3)}}(\omega_\text{THz})$. These results clarify the interactions in ZnTe at terahertz frequencies, with implications for measurements of intense terahertz fields using electro-optic sampling. 

\end{abstract*}

\section{Introduction}
The well-established technique of terahertz time-domain spectroscopy (THz-TDS) allows the simultaneous measurement of the magnitude and phase of a terahertz signal~\cite{koch_terahertz_2023} using for example the linear electro-optic effect of materials such as ZnTe in electro-optic sampling (EOS) detection~\cite{wu_free-space_1995, jepsen_fseos}. THz-TDS combined with EOS detection represents a powerful technique for measuring the complex refractive index at terahertz frequencies of samples placed in the terahertz beam and can also give insights about the material utilized in the EOS detection itself.
With high electric field terahertz pulses~\cite{hebling2008generation}, it is also possible to study the nonlinear behavior of different materials at terahertz frequencies. This has been shown in a wide range of studies, including terahertz-driven nonlinear spin control~\cite{baierl_nonlinear_2016}, terahertz high-harmonic generation by hot carriers~\cite{hafez_extremely_2018}, terahertz-induced ferroelectricity and collective coherence control of ferroelectric crystals~\cite{qi_collective_2009}. In addition, large nonlinear refractive indices at terahertz frequencies have been reported in crystals~\cite{Kerr_THz}; for example, Zibod \textit{et al.} showed a strong nonlinear response in crystalline quartz at terahertz frequencies~\cite{Kerr_THz2}.
Terahertz-induced nonlinear effects such as the Kerr effect at visible frequencies have also been reported for different liquids~\cite{freysz_terahertz_2010, tcypkin_giant_2021}, in amorphous chalcogenide glasses~\cite{zalkovskij_terahertz-induced_2013}, in common optical window and substrate materials~\cite{sajadi_terahertz-field-induced_2015}, and also in crystals like GaP and ZnTe~\cite{cornet_terahertz_2014, he_direct_2006}.

Due to the frequent use of ZnTe in EOS setups, it has been the subject of considerable study. Indeed, <110>-cut ZnTe is one of the most commonly used detection crystals in terahertz EOS setups. Several studies investigated the nonlinear effects at \textit{optical freqencies} in ZnTe with terahertz-pump optical-probe measurements~\cite{he_direct_2006,tian_quantitative_2008,zhukova_experimental_2017,chen_influence_2009,caumes_kerr-like_2002}. It has a zincblende structure and exhibits multiple nonlinear effects, such as second harmonic generation~\cite{wagner1998dispersion} and optical rectification~\cite{Planken:01}. It also has a strong Pockels effect~\cite{cornet2014terahertz}, and second-order nonlinear effects such as the Kerr effect can be observed~\cite{tian_quantitative_2008}.

However, so far there has been less focus on the investigation of terahertz-induced nonlinear processes in ZnTe at \textit{terahertz frequencies}, as would be revealed by, for example, a terahertz-pump, terahertz-probe measurement.

Cross \textit{et al.} investigated how optical and terahertz beams interact in ZnTe based on direct and cascaded nonlinear optical processes in ZnTe. They observe a cascading second-order nonlinear effect, based on a first-order nonlinear interaction of the terahertz pump and terahertz probe pulse followed by a second interaction of the resulting field with the optical probe pulse.
However, while noting that this first order nonlinear terahertz pump pulse - terahertz probe pulse interaction occurs, they focused their analysis solely on the cascading interactions with the optical beam at IR frequencies, mentioning that the latter are about five orders of magnitude stronger for cascading first order nonlinear effects~\cite{Martin_Cross}.

The occurrence of second order nonlinear processes was recently shown by Liu \textit{et al.} who reported electro-optic Kerr effects in 2-D terahertz spectroscopy~\cite{liu2024excitation}, but so far no study has performed a quantitative analysis of the nonlinear response. Given the ubiquity of ZnTe in EOS measurements, such nonlinear processes induced by intense terahertz radiation are of considerable relevance and interest.

Here, we report a terahertz-pump, terahertz-probe study of ZnTe, using a tilted-pulse-front source as the terahertz pump and a spintronic terahertz emitter as the terahertz probe, both pumped by the same femtosecond laser. Our setup enables easy spatial and temporal overlap of the two terahertz pulses.
We observe interactions between the terahertz pump and terahertz probe beam in ZnTe which reflect the nonlinearity of the material at terahertz frequencies. Analysis of these signals allows us to extract the nonlinear optical constants in the terahertz regime.

\section{Methods}
\begin{figure}[ht]
    \centering
    \includegraphics[scale = 1]{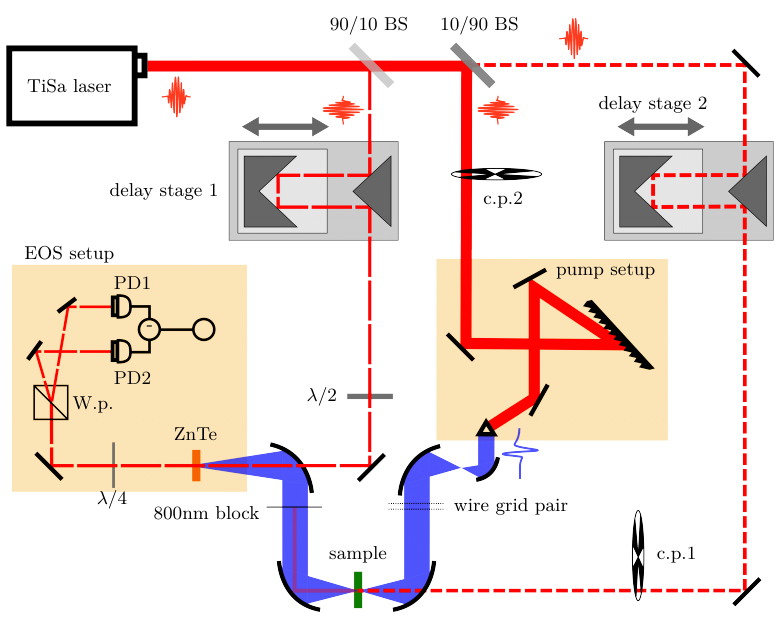}
    \caption{Experimental setup as it is employed in the experiments described in the following. TPF (terahertz pump) pump beam: thick solid line, detection beam: long-period dashed, STE (terahertz probe) pump beam: short-period dashed. Abbreviations: BS, beam splitter; c.p., chopper position; W.p., Wollaston prism; PD, photo diode.
    }
    \label{fig:exp_setup_STE_TPF}
\end{figure}

The experimental setup, shown in Fig.~\ref{fig:exp_setup_STE_TPF}, consists of a high-intensity tilted-pulse-front (TPF) terahertz beam~\cite{hebling2008generation} (terahertz pump) pumped by a strong optical pump beam (TPF pump, thick solid line), a spintronic terahertz emitter (STE) beam~\cite{kampfrath_terahertz_2013} (terahertz probe) pumped by a different optical beam (STE pump, short-period dashed), as well as an optical detection beam (optical detection, long-period dashed) for the electro-optic sampling (EOS) detection. All three optical beam paths originate from an amplified Ti:sapphire laser with a center wavelength of \SI{800}{nm}, a pulse length of \SI{80}{fs}, a repetition rate of \SI{1}{kHz} and a maximum output power of \SI{6}{W}. 

Most of the output power ($\sim$\SI{80}{\%}) of the femtosecond laser is used to pump a lithium niobate (LiNbO$_3$) crystal in the tilted-pulse-front configuration, to generate high intensity terahertz pulses~\cite{hebling2008generation,toth2023tilted,keiser2021structural}. The peak electric field of the terahertz pump as measured with a pyroelectric sensor (QMC instruments) and a terahertz camera (Terahertz Wave Imager, NEC) is \SI{100\pm10}{kV/cm} and can be attenuated with a pair of wiregrid polarizers.
Off-axis parabolic mirrors ($f=\SI{2}{in}$) create a focus of the terahertz pump beam, indicated as sample position in Fig.~\ref{fig:exp_setup_STE_TPF}. The terahertz pump is then refocused onto a $\langle110\rangle$-cut ZnTe crystal in the EOS setup. Meanwhile, $\sim$\SI{10}{\%} of the femtosecond laser output power is used for the optical detection beam in the EOS detection. Before reaching the ZnTe crystal, the optical detection beam passes through a delay stage (labeled delay stage 1 in Fig.~\ref{fig:exp_setup_STE_TPF}) to control the time delay between the optical detection pulse and the terahertz pump and terahertz probe pulses, and through a $\lambda/2$ wave plate to control its polarization. 
The remaining part of the femtosecond laser output power ($\sim$\SI{10}{\%}) also passes through a delay stage (labeled delay stage 2 in Fig.~\ref{fig:exp_setup_STE_TPF}) and is then focused onto a STE wafer located at the focus of the terahertz pump beam. The optical pump pulse is partially absorbed by the STE wafer which then generates the terahertz probe pulse. Transmission of terahertz radiation through the thin metal layers of the STE wafer is about \SI{70}{\%}, so the terahertz pump pulse from the TPF is only slightly attenuated after passing through the STE wafer. Because the probe pulse is emitted at the focus between the off-axis parabolic mirrors, the transmitted part of the optical STE pump pulse, the terahertz probe pulse, and the terahertz pump pulse all follow the same beam path afterward. The residual \SI{800}{nm} light is blocked by a sheet of high density polyethylene (transparent at terahertz frequencies), and the two remaining terahertz pulses are then focused onto the ZnTe crystal of the EOS setup. Delay stage 2 therefore controls the time delay between the terahertz pump and terahertz probe pulses in the ZnTe.

The STE for generating the terahertz probe pulse consists of a \linebreak W(\SI{2.0}{nm})$\big\vert$FeCoB(\SI{1.8}{nm})$\big\vert$Pt(\SI{2.0}{nm}) tri-layer structure, deposited onto \SI{0.5}{mm} thick sapphire substrates employing RF-diode sputtering, where the FeCoB is an alloy with the stoichiometry of $\mathrm{Fe}_{60}\mathrm{Co}_{20}\mathrm{B}_{20}$. Details about the manufacturing process can be found in~\cite{paries_fiber-tip_2023}.
Due to damage to the STE when conducting the measurements, multiple STE samples were employed during the course of the experiments. By manufacturing multiple STE wafers in one sputtering process, we ensure that the measurements with the different STEs are comparable~\cite{paries_optical_2024} (see also Fig. S1).

 All measurements are performed with the terahertz probe pulse polarized parallel to the terahertz pump pulse and the detection polarization along the [001] axis of the $\langle110\rangle$-cut ZnTe crystal. To increase the sensitivity of the EOS detection, we use a lock-in amplifier with a mechanical chopper at a frequency of \SI{743}{Hz}. To separate the terahertz probe signal from the terahertz pump signal, the chopper can be placed in either of the corresponding optical pump paths (labeled chopper positions 1 and 2 in Fig.~1). When the chopper is placed in the terahertz probe path, the terahertz pump beam is not modulated and the corresponding signal is not amplified.
    The intensities of the optical beams are controlled by placing filters in either of the beam paths.

\begin{figure}[htb]
    \centering
    \includegraphics[scale = 1]{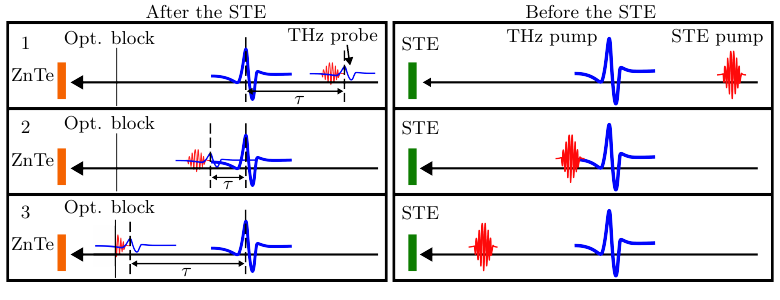}
    \caption{Schematic drawing of the relative time delay between the pulses at two different points (left and right panel) in the experiment for 3 different delay stage positions of the optical STE pump pulse (1-3). On the right-hand side, the respective pulses are displayed before reaching the STE and on the left-hand side the respective pulses are displayed after reaching the STE. The beams enter the STE  from the right side, to mirror the propagation direction of the beams and the geometry in Fig. 1. On the right-hand side, the terahertz pump pulse as well as the optical STE pump pulse are displayed. On the left-hand side the terahertz pump pulse amplitude as well as the optical STE pump pulse amplitude decrease due to absorption and reflection in the STE wafer. Also, the terahertz probe pulse is visible with a terahertz pulse amplitude smaller than the terahertz pump pulse amplitude. The relative time delay between the terahertz pump pulse and the terahertz probe pulse is marked as $\tau$.}
    \label{fig:exp_setup_STE_TPF_time_delay}
\end{figure}

To illustrate the measurement method, Fig.~\ref{fig:exp_setup_STE_TPF_time_delay} shows a schematic drawing of the relative time delay between the pulses. The right-hand side panels illustrate the terahertz pump and the optical STE pump pulses arriving at the STE, for 3 different delay stage positions of the optical STE pump pulse.  The left-hand side panels additionally show the terahertz probe pulse emitted by the STE for different relative time delays $\tau$ between the terahertz pump pulse and the terahertz probe pulse.

\section{Results and Discussion}

\begin{figure}[b]
    \centering
    \includegraphics[scale=1]{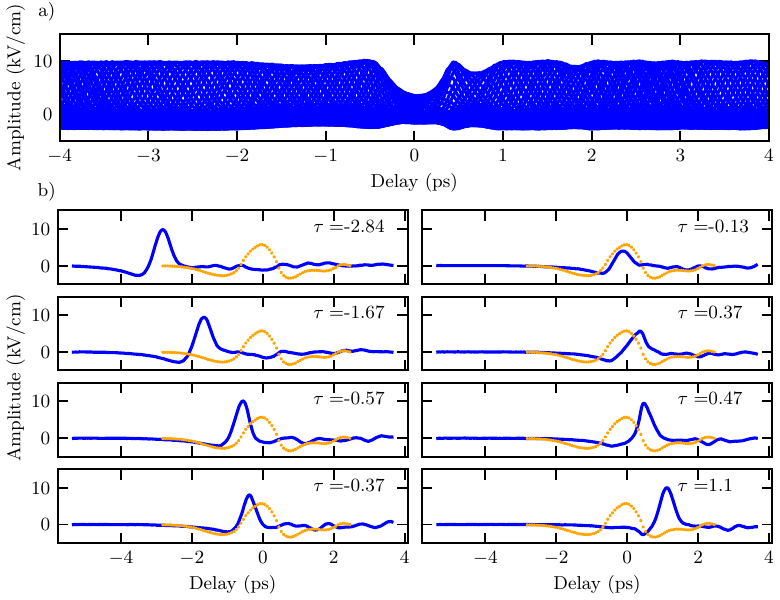}
    \caption{ a) Overlay of terahertz probe waveforms at different relative time delays $\tau$ between the terahertz pump pulse and the terahertz probe pulse. We observe a change in terahertz probe amplitude when temporally and spatially overlapping the terahertz probe pulse and the terahertz pump pulse.
    b) Selected terahertz probe waveforms at different relative time delays $\tau$ between the terahertz pump pulse and the terahertz probe pulse. The terahertz pump waveform (scaled by 0.06, orange dotted) is shown as a guide to the eye.}
    \label{fig:Kerr_observ}
\end{figure}

Figure~\ref{fig:Kerr_observ}a shows a series of terahertz probe waveforms overlaid for different time delays relative to the (fixed) terahertz pump. 
In these measurements, the TPF pulse (terahertz pump) has a peak electric field of \SI{100\pm10}{kV/cm}, while the STE pulse (terahertz probe) is approximately 10 times smaller. To ensure that the high peak electric field of the terahertz pump pulse does not influence the electro-optic response in ZnTe at optical frequencies by higher-order nonlinearities, we compared the electro-optic measurement of the terahertz waveform with the intensity measurement performed with the pyroelectric sensor. We found that the electro-optic response is linear up to our maximum peak electric field. Details can be found in the supplementary document (Fig. S2).

The optical chopper is in position 1, so that the measured signal is \textit{only} the terahertz probe, and the (much more intense) terahertz pump beam is not detected. Each of the waveforms displayed in Fig.~\ref{fig:Kerr_observ}a is measured by scanning delay stage 1 (the optical detection beam delay), with a fixed position of delay stage 2. 170 different positions of delay stage 2 are shown in Fig.~\ref{fig:Kerr_observ}a, superposing all of these waveforms in a single plot. The key feature of this data set is the variation of the peak amplitude of the probe pulse as a function of its delay relative to the pump pulse (which remains fixed in all measurements, and which defines the zero of the time axis in these plots). The terahertz probe amplitude clearly drops during the time when it is overlapped with the pump pulse in the ZnTe, and its amplitude is modulated by smaller amounts only for positive delays. This effect has a nonlinear connection to the pump pulse electric field.

Figure~\ref{fig:Kerr_observ}b shows eight of the terahertz probe waveforms from the same measurement, plotted individually for a better visualization of the effect. 
These plots also show the terahertz pump pulse waveform (dotted orange curves, measured separately by moving the chopper to position 2) to provide a visual reference of the relative time delay $\tau$ between the pump and probe pulses.

To clearly establish that this effect occurs in the ZnTe crystal and not in the STE layer itself, a silicon wafer was added after the STE wafer. If the effect originates in the STE layer, the ratio between the THz probe waveform amplitude and the strength of the effect should be independent of the subsequent attenuation: the signal measured after the wafer would depend linearly on the THz probe and THz pump amplitude. We observed a  nonlinear change in effect strength which proves that the observed effect is indeed localized in the ZnTe crystal. A more detailed description can be found in the supplementary document.

\subsection*{Model}

To explain the origin of the observed changes in the probe waveform, we model the measured data based on the idea of a terahertz-induced Kerr nonlinearity in the ZnTe detection crystal, in the terahertz frequency regime~\cite{liu2024excitation}. This contrasts with previous studies of terahertz-induced Kerr-type nonlinearities in the optical frequency range~\cite{Kerr_THz, Kerr_THz2}. The effect can be described using the dielectric polarization of the optical probe pulse according to the terahertz field-induced polarization in the ZnTe crystal.
The polarization in the ZnTe crystal, $P_\text{opt}(t)$, consists of both a linear term and nonlinear terms which are connected to the terahertz electric fields in ZnTe. High fields can induce polarization changes in ZnTe which affect the probe pulse\cite{liu2024excitation}.
For static, or low frequency, electric fields the induced nonlinear polarization is described by the electro-optic Pockels effect and the electro-optic Kerr effect~\cite{boyd_nonlinear_2019_chap11}. $P_\text{opt}(t)$ can therefore be expressed as:
\begin{equation}
\begin{split}
    P_\text{opt}(t) &= P_\text{opt, lin}^{(1)}(t) + P_\text{opt, Pockels}^{(2)}(t) + P_\text{opt, Kerr}^{(3)}(t) + ...\\ 
         &= \epsilon_{0}[\chi^{(1)}E_\text{opt}(t) + \chi^{(2)}E_\text{opt}(t)\ E_\text{THz}(t) + \chi^{(3)}E_\text{opt}(t)\ E_\text{THz}^2(t) + ...].
\end{split}
\label{eq:polar}
\end{equation}
where $P_\text{opt, lin}^{(1)}(t)$ describes the linear response of the polarization, $P_\text{opt, Pockels}^{(2)}(t)$ the linear electro-optic effect, $P_\text{opt, Kerr}^{(3)}(t)$ the quadratic electro-optic effect and the $\chi{(i)}$ the different orders of the susceptibility. $E_\text{THz}$ denotes a superposition of the two terahertz fields present: $E_\text{probe}$ and $E_\text{pump}$. Since $E_\text{probe}\ll E_\text{pump}$, we may assume $E_\text{THz}^2\approx E_\text{pump}^2$ in the following and aim to find an accurate expression for $E_\text{THz}$.

In previous terahertz-pump optical-probe studies, it was shown that second-order nonlinear interactions between an optical probe pulse and a terahertz pump pulse play only a minor role at \textit{optical} frequencies~\cite{caumes_kerr-like_2002, tian_quantitative_2008}. We therefore neglect $P_\text{opt, Kerr}^{(3)}(t)$, simplifying Eq.~\eqref{eq:polar} to:
\begin{equation}
\begin{split}
    P_\text{opt}(t) &= P_\text{opt, lin}^{(1)}(t) + P_\text{opt, Pockels}^{(2)}(t)\\ 
         &= \epsilon_{0}[\chi^{(1)}E_\text{opt}(t) + \chi^{(2)}E_\text{opt}(t)\  E_\text{THz}(t)].
\end{split}
\label{eq:polarShort}
\end{equation}

To find the right expression for $E_\text{THz}$ in $P_\text{opt}(t)$, we are now looking into the polarization $P_\text{probe}(t)$ which the terahertz pulses experience in the crystal considering the interaction between the terahertz probe pulse and the terahertz pump pulse in the ZnTe crystal~\cite{boyd_nonlinear_2019_chap11}.
It was shown in~\cite{Kerr_THz} and~\cite{Kerr_THz2} that crystals can exhibit a strong second-order nonlinear response in the terahertz range due to phonon resonances. Therefore, in contrast to the assumption in Eq. \eqref{eq:polarShort}, we cannot neglect the second-order nonlinear response at terahertz frequencies, $P_\text{THz, Kerr}^{(3)}(t)$.

Because $E_\text{probe}\ll E_\text{pump}$, we neglect terms that are quadratic in $E_\text{probe}$. 
These assumptions result in:
\begin{equation}
\begin{split}
    P_\text{probe}(t) &= \epsilon_{0}[\chi^{(1)}E_\text{probe}(t) + \chi^{(1)}E_\text{pump}(t) + 2\  \chi^{(2)}E_\text{probe}(t)\  E_\text{pump}(t)  \\
    &+ 3\  \chi^{(3)}E_\text{probe}(t)\  E_\text{pump}^2(t)],\\
\end{split}
\label{eq:polarTHz}
\end{equation}
By inserting the E-field resulting from Eq.~\eqref{eq:polarTHz} as $E_\text{THz}(t)$ in Eq.~\eqref{eq:polarShort}, we find:
\begin{equation}
\begin{split}
    P_\text{opt}(t) &\propto  \epsilon_{0}\bigl\{\chi^{(1)}E_\text{opt}(t) + \chi^{(2)}E_\text{opt}(t)\  [A\  E_\text{probe}(t) + B\  E_\text{pump}(t)\\
    &+ C\  E_\text{probe}(t)\  E_\text{pump}(t) + D\  E_\text{probe}(t)\  E_\text{pump}^2(t)]\bigr\}
\end{split}
\label{eq:polarlong}
\end{equation}

In the experiments, the optical STE pump pulse is modulated by the chopper. Therefore, only signals modulated with $E_\text{probe}(t)$ are detected, and terms in Eq.~\eqref{eq:polarlong} that do not contain $E_\text{probe}(t)$ can be neglected. This results in:
\begin{equation}
\begin{split}
    P_\text{opt}(t) &\propto  \epsilon_{0}\bigl\{\chi^{(2)}E_\text{opt}(t)\  [A\  E_\text{probe}(t)\\
    &+ C\  E_\text{probe}(t)\  E_\text{pump}(t) + D\  E_\text{probe}(t)\  E_\text{pump}^2(t)]\bigr\}
\end{split}
\label{eq:polarfinal}
\end{equation}

To model the entire signal, the change in transmittance of the STE wafer after optical excitation has to be taken into account. A small part of the terahertz pump signal passing through the wafer is changed by the  change in transmittance of the STE wafer due to the optical excitation, which is itself modulated by the chopper. This change in the terahertz pump is also detected by the lock-in amplifier.
This signal can be written as $\epsilon_0 \chi^{(1)} \Delta E_\text{pump}(t) = \epsilon_0 \chi^{(1)} \tilde{b} E_\text{pump}(t)$ with $\tilde{b}<1$. Adding this to the signal results in the final model:
\begin{equation}
\begin{split}
    P_\text{opt}(t) &\propto  \epsilon_{0}\bigl\{\chi^{(2)}E_\text{opt}(t)\  [A\  E_\text{probe}(t) + \tilde{b}\  E_\text{pump}(t)\\
    &+ C\  E_\text{probe}(t)\  E_\text{pump}(t) + D\  E_\text{probe}(t)\  E_\text{pump}^2(t)]\bigr\}
\end{split}
\label{eq:polarfinalTPF}
\end{equation}

When fitting this model to data, we also have to take causality into account. Specifically, the probe pulse cannot interact with portions of the pump pulse that arrive at the ZnTe crystal before the probe. We therefore also multiply with a half-window Heaviside function $\theta(t-t_0)$ where $t_0$ denotes the time at which the pump pulse first arrives at the sample position. 

After combining the different prefactors in equation~\eqref{eq:polarfinalTPF}, the following equation is used for the model:
\begin{equation}
\begin{split}
    E_\text{model}(t) &=  E_\text{probe}(t) + \theta(t-t_0) \  \bigl[ a\   E_\text{pump}(t)+ b\  E_\text{probe}(t)\  E_\text{pump}(t) \\ 
    &+ c\  E_\text{probe}(t)\  E_\text{pump}^2(t)\bigr]
\end{split}
\label{eq:model}
\end{equation}

\begin{figure}[h]
    \centering
    \includegraphics[scale=1]{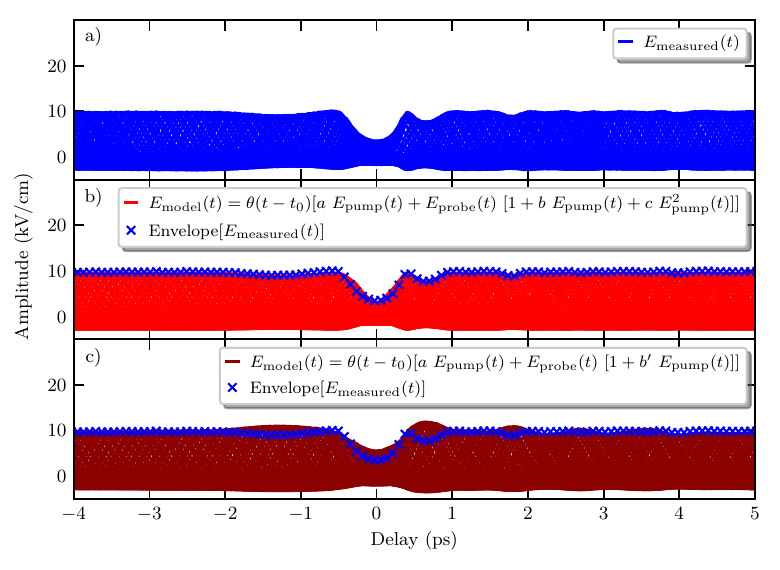}
    \vspace{-3mm}
    \caption{Comparison of the terahertz pump induced change in terahertz probe amplitude. a) Overlay of the measured terahertz probe waveforms at different relative time delays between the terahertz pump pulse and the terahertz probe pulse. The entirety of the waveforms is here described by $E_\text{measured}(t)$. b) Fitted model according to Eq. (7). The fit parameters are: $a=0.013\pm0.002$, $b=(3.6\pm1.7)10^{-9}\text{m}/\text{V}$, and $c=(6.7\pm0.7)10^{-15}\text{m}^{2}/\text{V}^2$. c) An equivalent fit, except with the parameter c constrained to be equal to zero ($a=0.013\pm0.002$ and $b^{'} =(4.4\pm0.5)10^{-8}\text{m}/\text{V}$), resulting in a fit with inferior quality. The blue crosses are representing the peak of the envelope of $E_\text{measured}(t)$.
    }
    \label{fig:MeasurementVSModel}
\end{figure}

After acquiring the reference waveforms $E_\text{probe}(t)$ of the terahertz probe pulse and $E_\text{pump}(t)$ of the terahertz pump pulse, the model is fitted to the experimental data. The results are shown in Fig.~\ref{fig:MeasurementVSModel}. For better visualization of the accuracy of the fit, the envelope of the measured waveforms is represented as blue crosses in Fig.~\ref{fig:MeasurementVSModel}b and c. The model captures the main features of the experimental data very well. The terahertz interactions in ZnTe can therefore be described as a terahertz-induced electro-optic Kerr nonlinearity at terahertz frequencies induced by the terahertz pump pulse and probed by the terahertz probe pulse.

We obtain the fit parameters $a=0.013\pm0.002$, $b=(3.6\pm1.7)10^{-9}\text{m}/\text{V}$, and $c=(6.7\pm0.7)10^{-15}\text{m}^{2}/\text{V}^2$. Hereby, $a$ describes the change in transmittance of the STE wafer due to the optical excitation, $b/2$ is an estimation for the second order susceptibility  $\chi^{(2)}\approx 1.8\times 10^{-9}\text{m}/\text{V}$ and $c/3$ an estimation for the third order susceptibility $\chi^{(3)}_\text{fit}\approx 2.2\times10^{-15}\text{m}^{2}/\text{V}^2$.

Wagner \textit{et al.} reported  values for $\chi^{(2)}$ at optical frequencies up to $\chi^{(2)}\approx0.45\times10^{-9}\text{m}/\text{V}$, which is of the same order of magnitude as our obtained value~\cite{wagner1998dispersion}.

We further emphasize that it is not possible to describe our results without including the term proportional to $E_\text{pump}^2(t)$ in Eq.~\eqref{eq:model}. Figure 4c demonstrates this fact, showing the best fit to the data with the parameter $c$ constrained to zero, producing a clearly inferior fit.

To further support the model of a terahertz-induced Kerr nonlinearity in the terahertz frequency regime, we measured the strength of the effect for different terahertz pump electric field amplitudes. 
The effect strength is characterized by a decrease in amplitude of the probe pulse when the pump pulse is present compared to when it is not present. For different pump-probe time delays, we therefore add the induced intensity differences:
\begin{equation}
    P_\text{Kerr}= \sum_{i=0}^N \mathrm{max}(V_i^{\mathrm{Ref}}) - \mathrm{max}(V_i^{\mathrm{Measured}}),
\label{eq:TPF_strength}
\end{equation}
where $V_i^{\mathrm{Ref}}$ is the $i^{th}$ measured reference terahertz probe waveform without the terahertz pump pulse, $V_i^{\mathrm{Measured}}$ is the $i^{th}$ measured reference terahertz probe waveform with the terahertz pump pulse present and $N$ is the number of waveforms taken in one measurement.

\begin{figure}[htb]
    \centering
    \includegraphics[scale=1]{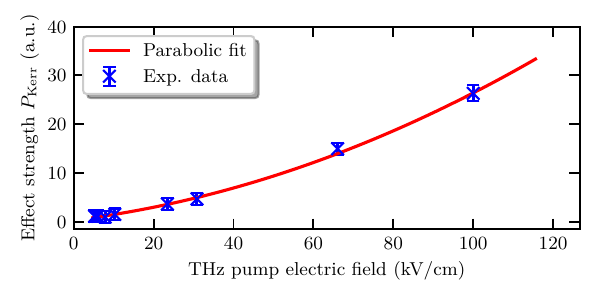}
    \caption{Calculated effect strength of the observed change in terahertz probe amplitude according to Eq.~\eqref{eq:TPF_strength}, plotted as a function of the terahertz pump electric field strength. The points are fitted with a parabolic function. The error bars are calculated from the standard
deviation of all points $[\mathrm{max}(V_i^{\mathrm{Ref}}) - \mathrm{max}(V_i^{\mathrm{Measured}})]$ of Eq.~\eqref{eq:TPF_strength}.}
    \label{fig:Kerr_intensity_dependence}
\end{figure}

In Fig.~\ref{fig:Kerr_intensity_dependence}, we plot the strength of the effect $P_\text{Kerr}$ versus the amplitude of the terahertz pump electric field.
The results confirm the expected quadratic relationship as predicted by Eq.~\eqref{eq:model}.

We note that such a nonlinear response is  important to consider for all terahertz-pump - terahertz-probe experiments using a common EOS setup with a ZnTe crystal and should not be neglected especially in experiments involving high intensity terahertz fields.

\subsection*{Estimating nonlinear coefficients} 
The uncertainty in the values of the electric field influences the absolute value of $\chi^{(3)}_\text{fit}$. However, we found two more ways to estimate $\chi^{(3)}$ from measurements. We first calculate  $\chi^{(3)}_\text{phase shift}$ from the observed decrease in probe pulse amplitude when the pump and probe pulses are overlapping (see Fig.\ref{fig:Kerr_observ}). 
We then also estimate $\chi^{(3)}_\text{time shift}$ using a different measurement, namely by tracking the field-dependent shift in the time-domain waveform as a result of the change in nonlinear refractive index.

\subsubsection*{A. Calculation from Phase Shift} \label{Sec:amp_dec}

The observed change in amplitude of the terahertz probe beam can be described as a phase change between the two polarization axes in the EOS detection $\Delta \varphi$ with $A_\text{min} = A_\text{max}\ \cos{(\Delta \varphi)}$. We estimate the amplitude decrease in Fig.~\ref{fig:Kerr_observ}b to be \SI{60\pm4}{\percent}. This results in a phase change of $\Delta \varphi = 1.15\pm0.05$. Details of the derivation can be found in the Supplementary Information. With 
\begin{equation}
    \Delta \varphi = \frac{2\pi}{\lambda_0}d\Delta n \text{\qquad\cite{demtroder_elektrizitat_2013}}
\end{equation}
and
\begin{equation}
    \chi^{\mathrm{(3)}} = \frac{2 \Delta n n_0}{3 \left|E\right|^2}\text{\qquad~\cite{boyd_nonlinear_2019_chap1}},
    \label{eq:Chi3}
\end{equation}
we calculate the third order susceptibility $\chi^{(3)}_\text{phase shift}$:
\begin{equation}
    \chi^{\mathrm{(3)}}_\text{phase shift} = \frac{2 n_0}{3 \left|E\right|^2}\frac{\Delta \varphi \lambda_0}{2\pi d} = (1.94\pm0.85)10^{\mathrm{-16}} \frac{\mathrm{m^2}}{\mathrm{V^2}}    
\end{equation}
Here $n_0=3.18$ is the ordinary, weak-field refractive index of ZnTe at \SI{1}{THz}~\cite{tripathi_accurate_2013}, \break$E=\SI{100\pm10}{kV/cm}$ is the measured peak electric field in the experiment, $d=\SI{2.00\pm0.01}{mm}$ is the thickness of the ZnTe crystal and $\lambda_0=\frac{c}{f n} =  \SI{100\pm20}{\micro m}$ is the estimated center wavelength at \SI{1}{THz} in ZnTe.

\subsubsection*{B. Calculation from Time Shift} \label{Sec:time_shift}
The second order non-linearity in the ZnTe crystal also leads to a change in refractive index. This change of the refractive index influences the time delay of the terahertz pump pulse inside the ZnTe crystal, which we expect to see in a time shift in the measured waveform that correlates with the terahertz peak amplitude. This provides an independent method to verify the previously obtained value $\chi^{\mathrm{(3)}}_\text{phase shift}$ which was measured at a fixed terahertz pump field strength.

For these field strength-dependent measurements, an additional ZnTe crystal is placed at the sample position indicated in Fig.~1 (instead of the STE wafer) and the terahertz pump waveform is acquired for different terahertz peak amplitudes. These measurements were performed in a nitrogen-purged environment to rule out any nonlinear effects due to interactions with ambient air, as discussed in~\cite{rasekh_terahertz_2021}. The results are shown in Fig.~6a, where a clear shift to higher time delays is visible in the waveforms. While timing laser jitter might also result in a time shift, we do not believe that laser jitter alone is responsible for the observed offset because the results are repeatable and the waveforms in Fig.~6 were acquired over multiple minutes, so that random laser jitter averages out.

\begin{figure}[ht]
    \centering
    \includegraphics[scale=1]{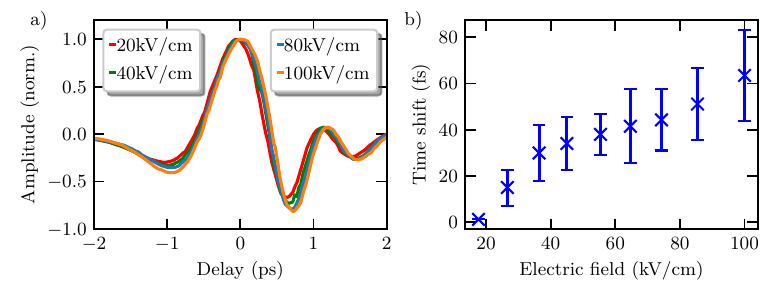}
    \caption{a) Selected terahertz pump pulses scaled to the lowest intensity pulse. With increasing intensity, a right shift is observed. b) Calculated time shift for different terahertz pump electric field peak powers. The error bars are calculated from the standard deviation of the time shifts $t(V_k^i)-t(V^{\mathrm{low}}=V_k^i)$ for all data points. 
    }
    \label{fig:time_shift}
\end{figure}

To quantify the value of $\chi^{\mathrm{(3)}}_\text{time shift}$, the time shift is calculated following the method used by Zibod \textit{et al.}~\cite{Kerr_THz2}:
\begin{equation}
    t_{av}^i=\frac{1}{N}\sum_k^N t(V_k^i)-t(V^{\mathrm{low}}=V_k^i).
\end{equation}
where $t_{av}^i$ is the average time shift for the measurement with the terahertz peak amplitude level $i$, $N$ is the number of data points in one measurement, $V^i$ the measured signal of terahertz peak amplitude level $i$, and $V^{\mathrm{low}}$ is the measured signal of the lowest terahertz peak amplitude. The calculated time shift is shown in Fig.~\ref{fig:time_shift}b. An approximately linear behavior is observed, with a maximum time shift of around \SI{63\pm19}{fs} at the largest terahertz peak electric field.

With Eq.~\eqref{eq:Chi3} and using
\begin{equation}
    \Delta n = \frac{c\Delta t}{d},
\end{equation}
we can calculate $\chi^{\mathrm{(3)}}_\text{time shift}$ as follows:
\begin{equation}
    \chi^{\mathrm{(3)}}_\text{time shift} = \frac{2 n_0}{3 \left|E\right|^2}\frac{c\Delta t}{d} = (2.0\pm1.0)10^{\mathrm{-16}} \frac{\mathrm{m^2}}{\mathrm{V^2}}    
\end{equation}
This result agrees very well with $\chi^{\mathrm{(3)}}_\text{phase shift}$ calculated using the phase change in the polarization.

We can now compare the three values for $\chi^{(3)}$: $\chi^{(3)}_\text{fit}=(22\pm2)10^{-16}\text{m}^{2}/\text{V}^2$, 
$\chi^{(3)}_\text{phase shift}=(1.9\pm0.9)10^{-16}\text{m}^{2}/\text{V}^2$, and  
$\chi^{(3)}_\text{time shift}=(2\pm1)10^{-16}\text{m}^{2}/\text{V}^2$.

We observe that different measurement and analysis methods lead to slightly different results. A visualization of the impact of the different results for $\chi^{(3)}$ can be found in Fig.~S4.
While the largest uncertainty is surely the value of the electric field, we also note that $\chi^{(3)}_\text{phase shift}$ was calculated at a single center frequency, hence excluding dispersion effects. Additionally, the discrepancy could partially arise from the before mentioned effects in the STE wafer itself due to the presence of the THz pump acting as a bias field. Even though the main observed second-order nonlinear effect was clearly localized in the ZnTe crystal, the interaction in the STE wafer could still lead to some kind of smaller effect which could contribute to the discrepancy between the measurements. We also only used two values from the pump-probe experiment (the probe field without the pump field present, and the probe field during overlap with the pump field). In this regard, we expect $\chi^{(3)}_\text{fit}$ to be a more accurate estimate: it includes dispersion effects and covers the entire pump-probe delay range.
$\chi^{(3)}_\text{time shift}$ was calculated from a different data set and it also includes dispersion effects, but it depends strongly on the strength of the pump pulse.

We conclude that $\chi^{(3)}$ of ZnTe in the terahertz range is on the order of $10^{-16}-10^{-15}\text{m}^{2}/\text{V}^2$. 
This result is of the same order of magnitude as predicted for other crystals such as crystalline quartz~\cite{Kerr_THz} and several orders of magnitude higher than the corresponding values for such materials at optical frequencies~\cite{caumes_kerr-like_2002, tian_quantitative_2008}.

\section{Conclusion}
In this work, we show a quantitative analysis of a second-order nonlinear effect in ZnTe at terahertz frequencies. With a terahertz-pump terahertz-probe setup, we measured the polarization change in ZnTe due to a high-intensity terahertz field. Additionally, we developed a model for the polarization which is in good agreement with the experiments and explains the interaction between the different electromagnetic pulses in the ZnTe crystal. Lastly, we calculated the strength of the nonlinear effect by analyzing both the polarization change in the ZnTe crystal and the time shift of a high intensity terahertz pulse inside the ZnTe crystal. Both methods are in good agreement and result in a third-order susceptibility $\chi^{(3)}$ on the order of $10^{-16}-10^{-15}\text{m}^{2}/\text{V}^2$. To the best of our knowledge, such a quantitative analysis of the terahertz nonlinearity of ZnTe has not been performed before. These results will inform all terahertz-pump terahertz-probe experiments using the most commonly employed EOS setup.

\section{Backmatter}

\begin{backmatter}
\bmsection{Funding} NSF ECCS-2300152, Deutsche Forschungsgemeinschaft (DFG, German Research Foundation) - TRR 173 - 268565370 (SPIN+X, project B11), Study Abroad Studentship of the German Academic Scholarship Foundation (Studienstiftung des deutschen Volkes)

\bmsection{Disclosures}
The authors declare no conflicts of interest.

\bmsection{Data availability} Data underlying the results presented in this paper are not publicly available at this time but may be obtained from the authors upon reasonable request.

\bmsection{Supplemental document}
See Supplement 1 for supporting content.

\end{backmatter}


\begin{thebibliography}{10}
\newcommand{\enquote}[1]{``#1''}

\bibitem{koch_terahertz_2023}
M.~Koch, D.~M. Mittleman, J.~Ornik, and E.~Castro-Camus, \enquote{Terahertz time-domain spectroscopy,} {\protect\JournalTitle{Nature Reviews Methods Primers}} \textbf{3}, 48 (2023).

\bibitem{wu_free-space_1995}
Q.~Wu and X.-C. Zhang, \enquote{Free-space electro-optic sampling of terahertz beams,} {\protect\JournalTitle{Applied Physics Letters}} \textbf{67}, 3523--3525 (1995).

\bibitem{jepsen_fseos}
P.~U. Jepsen, C.~Winnewisser, M.~Schall, \emph{et~al.}, \enquote{Detection of {THz} pulses by phase retardation in lithium tantalate,} {\protect\JournalTitle{Phy. Rev. E}} \textbf{53}, R3052--R3054 (1996).

\bibitem{hebling2008generation}
J.~Hebling, K.-L. Yeh, M.~C. Hoffmann, \emph{et~al.}, \enquote{Generation of high-power terahertz pulses by tilted-pulse-front excitation and their application possibilities,} {\protect\JournalTitle{JOSA B}} \textbf{25}, B6--B19 (2008).

\bibitem{baierl_nonlinear_2016}
S.~Baierl, M.~Hohenleutner, T.~Kampfrath, \emph{et~al.}, \enquote{Nonlinear spin control by terahertz-driven anisotropy fields,} {\protect\JournalTitle{Nature Photonics}} \textbf{10}, 715--718 (2016).

\bibitem{hafez_extremely_2018}
H.~A. Hafez, S.~Kovalev, J.-C. Deinert, \emph{et~al.}, \enquote{Extremely efficient terahertz high-harmonic generation in graphene by hot {Dirac} fermions,} {\protect\JournalTitle{Nature}} \textbf{561}, 507--511 (2018).

\bibitem{qi_collective_2009}
T.~Qi, Y.-H. Shin, K.-L. Yeh, \emph{et~al.}, \enquote{Collective coherent control: {Synchronization} of polarization in ferroelectric $\mathrm{PbTiO_3}$ by shaped {THz} fields,} {\protect\JournalTitle{Physical Review Letters}} \textbf{102}, 247603 (2009).

\bibitem{Kerr_THz}
K.~Dolgaleva, D.~V. Materikina, R.~W. Boyd, and S.~A. Kozlov, \enquote{Prediction of an extremely large nonlinear refractive index for crystals at terahertz frequencies,} {\protect\JournalTitle{Physical Review A}} \textbf{92}, 023809 (2015).

\bibitem{Kerr_THz2}
S.~Zibod, P.~Rasekh, M.~Yildrim, \emph{et~al.}, \enquote{Strong nonlinear response in crystalline quartz at {THz} frequencies,} {\protect\JournalTitle{Advanced Optical Materials}} \textbf{11}, 2202343 (2023).

\bibitem{freysz_terahertz_2010}
E.~Freysz and J.~Degert, \enquote{Terahertz {Kerr} effect,} {\protect\JournalTitle{Nature Photonics}} \textbf{4}, 131--132 (2010).

\bibitem{tcypkin_giant_2021}
A.~Tcypkin, M.~Zhukova, M.~Melnik, \emph{et~al.}, \enquote{Giant third-order nonlinear response of liquids at terahertz frequencies,} {\protect\JournalTitle{Physical Review Applied}} \textbf{15}, 054009 (2021).

\bibitem{zalkovskij_terahertz-induced_2013}
M.~Zalkovskij, A.~C. Strikwerda, K.~Iwaszczuk, \emph{et~al.}, \enquote{Terahertz-induced {Kerr} effect in amorphous chalcogenide glasses,} {\protect\JournalTitle{Applied Physics Letters}} \textbf{103}, 221102 (2013).

\bibitem{sajadi_terahertz-field-induced_2015}
M.~Sajadi, M.~Wolf, and T.~Kampfrath, \enquote{Terahertz-field-induced optical birefringence in common window and substrate materials,} {\protect\JournalTitle{Optics Express}} \textbf{23}, 28985 (2015).

\bibitem{cornet_terahertz_2014}
M.~Cornet, J.~Degert, E.~Abraham, and E.~Freysz, \enquote{Terahertz {Kerr} effect in gallium phosphide crystal,} {\protect\JournalTitle{Journal of the Optical Society of America B}} \textbf{31}, 1648 (2014).

\bibitem{he_direct_2006}
W.-Q. He, C.-M. Gu, and W.-Z. Shen, \enquote{Direct evidence of {Kerr}-like nonlinearity by femtosecond {Z}-scan technique,} {\protect\JournalTitle{Optics Express}} \textbf{14}, 5476 (2006).

\bibitem{wagner1998dispersion}
H.~P. Wagner, M.~K{\"u}hnelt, W.~Langbein, and J.~M. Hvam, \enquote{Dispersion of the second-order nonlinear susceptibility in {ZnTe}, {ZnSe}, and {ZnS},} {\protect\JournalTitle{Physical Review B}} \textbf{58}, 10494--10501 (1998).

\bibitem{Planken:01}
P.~C.~M. Planken, H.-K. Nienhuys, H.~J. Bakker, and T.~Wenckebach, \enquote{Measurement and calculation of the orientation dependence of terahertz pulse detection in ZnTe,} {\protect\JournalTitle{J. Opt. Soc. Am. B}} \textbf{18}, 313--317 (2001).

\bibitem{cornet2014terahertz}
M.~Cornet, J.~Degert, E.~Abraham, and E.~Freysz, \enquote{Terahertz-field-induced second harmonic generation through Pockels effect in zinc telluride crystal,} {\protect\JournalTitle{Optics Letters}} \textbf{39}, 5921--5924 (2014).

\bibitem{tian_quantitative_2008}
Z.~Tian, C.~Wang, Q.~Xing, \emph{et~al.}, \enquote{Quantitative analysis of {Kerr} nonlinearity and {Kerr}-like nonlinearity induced via terahertz generation in {ZnTe},} {\protect\JournalTitle{Applied Physics Letters}} \textbf{92}, 041106 (2008).

\bibitem{Martin_Cross}
M.~Cross, \enquote{Advances in two-dimensional {THz} spectroscopy: Distinguishing cascaded nonlinear optical processes in {ZnTe},} Ph.D. thesis, Technical University of Denmark (2023), p.79.

\bibitem{zhukova_experimental_2017}
M.~Zhukova, E.~Makarov, S.~Putilin, \emph{et~al.}, \enquote{Experimental study of {THz} electro-optical sampling crystals {ZnSe}, {ZnTe} and {GaP},} {\protect\JournalTitle{Journal of Physics: Conference Series}} \textbf{917}, 062021 (2017).

\bibitem{chen_influence_2009}
X.~Chen, S.~He, Z.~Shen, \emph{et~al.}, \enquote{Influence of nonlinear effects in {ZnTe} on generation and detection of terahertz waves,} {\protect\JournalTitle{Journal of Applied Physics}} \textbf{105}, 023106 (2009).

\bibitem{caumes_kerr-like_2002}
J.-P. Caumes, L.~Videau, C.~Rouyer, and E.~Freysz, \enquote{{Kerr}-like nonlinearity induced via terahertz generation and the electro-optical effect in zinc blende crystals,} {\protect\JournalTitle{Physical Review Letters}} \textbf{89}, 047401 (2002).

\bibitem{liu2024excitation}
A.~Liu and A.~Disa, \enquote{Excitation-dependent features and artifacts in 2-D terahertz spectroscopy,} {\protect\JournalTitle{Optics Express}} \textbf{32}, 28160--28168 (2024).

\bibitem{kampfrath_terahertz_2013}
T.~Kampfrath, M.~Battiato, P.~Maldonado, \emph{et~al.}, \enquote{Terahertz spin current pulses controlled by magnetic heterostructures,} {\protect\JournalTitle{Nature Nanotechnology}} \textbf{8}, 256--260 (2013).

\bibitem{toth2023tilted}
G.~T{\'o}th, G.~Pol{\'o}nyi, and J.~Hebling, \enquote{Tilted pulse front pumping techniques for efficient terahertz pulse generation,} {\protect\JournalTitle{Light: Science \& Applications}} \textbf{12}, 256 (2023).

\bibitem{keiser2021structural}
G.~R. Keiser, N.~Karl, S.~R. Ul~Haque, \emph{et~al.}, \enquote{Structural tuning of nonlinear terahertz metamaterials using broadside coupled split ring resonators,} {\protect\JournalTitle{AIP Advances}} \textbf{11}, 095103 (2021).

\bibitem{paries_fiber-tip_2023}
F.~Paries, N.~Tiercelin, G.~Lezier, \emph{et~al.}, \enquote{Fiber-tip spintronic terahertz emitters,} {\protect\JournalTitle{Optics Express}} \textbf{31}, 30884 (2023).

\bibitem{paries_optical_2024}
F.~Paries, F.~Selz, C.~N. Santos, \emph{et~al.}, \enquote{Optical damage thresholds of single-mode fiber-tip spintronic terahertz emitters,} {\protect\JournalTitle{Optics Express}} \textbf{32}, 24826 (2024).

\bibitem{boyd_nonlinear_2019_chap11}
R.~W. Boyd, \emph{Nonlinear optics}, 4th ed. (Academic Press, San Diego, 2019), Chap. 11.

\bibitem{demtroder_elektrizitat_2013}
W.~Demtröder, \emph{Electrodynamics and Optics}, 1st ed. (Springer Nature, Cham, Switzerland, 2019), Chap. 8.

\bibitem{boyd_nonlinear_2019_chap1}
R.~W. Boyd, \emph{Nonlinear optics}, 4th ed. (Academic Press, San Diego, 2019), Chap. 1.

\bibitem{tripathi_accurate_2013}
S.~R. Tripathi, M.~Aoki, M.~Takeda, \emph{et~al.}, \enquote{Accurate complex refractive index with standard deviation of {ZnTe} measured by terahertz time domain spectroscopy,} {\protect\JournalTitle{Japanese Journal of Applied Physics}} \textbf{52}, 042401 (2013).

\bibitem{rasekh_terahertz_2021}
P.~Rasekh, A.~Safari, M.~Yildirim, \emph{et~al.}, \enquote{Terahertz nonlinear spectroscopy of water vapor,} {\protect\JournalTitle{ACS Photonics}} \textbf{8}, 1683--1688 (2021).

\end{thebibliography}

\end{document}